\documentclass[aps,twocolumn,showkeys,amsmath,amssymb,floatfix,pre,superscriptaddress]{revtex4-1}
\usepackage{braket}
\usepackage{graphicx}
\usepackage{mathcomp}

\newcommand{\be}{\begin{equation}}
\newcommand{\ee}{\end{equation}}

\newcommand{\bea}{\begin{eqnarray}}
\newcommand{\eea}{\end{eqnarray}}

\begin{document}

\author{Eduardo Os\'orio Rizzatti}
\affiliation{Instituto de F\'isica, Universidade Federal de do Rio Grande do Sul, Porto Alegre-RS, Brazil}
\title{Waterlike anomalies on the Bose-Hubbard Model}
\author{M\'arcio Sampaio Gomes Filho}
\affiliation{Instituto de F\'isica, Universidade de Bras\'ilia, Bras\'ilia-DF, Brazil}
\author{Mariana Malard Sales}
\affiliation{Programa de P\'os-Gradua\c{c}\~ao em Ci\^encia de Materiais, Faculdade UnB Planaltina, Universidade de Bras\'ilia, Planaltina-DF, Brazil}
\affiliation{Department of Physics, University of Gothenburg, SE 412 96 Gothenburg, Sweden}
\author{Marco Aur\'elio A. Barbosa}
\email{aureliobarbosa@unb.br}
\thanks{Corresponding author}
\affiliation{Programa de P\'os-Gradua\c{c}\~ao em Ci\^encia de Materiais, Faculdade UnB Planaltina, Universidade de Bras\'ilia, Planaltina-DF, Brazil}

\date{\today}
 
\begin{abstract}
Although well-researched as a prototype Hamiltonian for strongly interacting quantum systems, the Bose-Hubbard model has not so far been explored as a fluid system with waterlike anomalies. In this work we show that this model supports, in the limit of a strongly localizing confining potential, density anomalies which can be traced back to ground state (zero-temperature) phase transitions between different Mott insulators. This key finding opens a new pathway for theoretical and experimental studies of liquid water and, in particular, we propose a test of our predictions that can be readily implemented in a ultra-cold atom platform. 
\end{abstract}


\maketitle

\section{Introduction}
The Hubbard model~\cite{Hubbard1963ElectronBands,Gutzwiller1963} is of great interest in many areas of condensed matter physics and has been extensively investigated through a variety of methods for strongly interacting quantum systems~\cite{Jaksch2005,Dutta2015}. In particular, the Bose-Hubbard model~\cite{fisher1989:prb,Sheshadri1993SuperfluidRPA,Freericks1994,Jaksch1998,Krutitsky2016} regained attention since its realization with cold bosonic atoms trapped on optical lattices~\cite{greiner2002:nature,Bloch2004QuantumLattices,Bloch2005,Windpassinger2013}. Indeed, such systems became a remarkable experimental arena for testing a myriad of theoretical concepts, playing the celebrated role of quantum simulators~\cite{Lewenstein2007UltracoldBeyond,Bloch2012QuantumGases}. \\
\indent In parallel, water is relevant for many reasons including its abundance on Earth, its role on chemistry of life and as a human resource~\cite{Brini2017,Franks2000Water:Life}. It also possess particular physicochemical properties, including its high latent heat, diffusion and thermal response functions~\cite{Brini2017,debenedetti04:review, Eisenberg2005, charusita2013:pccp:review, marcia:jcp, Gallo2016}. A striking property of water is the increase of density with temperature in the range from $0^\circ$C to $4^\circ$C, setting it apart from regular liquids~\cite{Eisenberg2005}. 
In liquid water, the temperature of maximum density (TMD) decreases with pressure entering the metastable regime above $40$ MPa~\cite{Angell1976, Caldwell1978}, and is associated to a region with negative value of the thermal expansion coefficient, $\alpha$. According to the second critical point (SCP) hypothesis the high temperature thermodynamic and dynamic anomalous behavior of liquid water is attributed to the presence of a metastable liquid-liquid phase transition ending in a critical point~\cite{poole92:nat,stanley08:epj}. The SCP hypothesis has been proposed from the observation of a liquid-liquid phase transition on computer simulations of the ST2 atomically detailed model of water~\cite{poole92:nat}, and was followed by extensive investigations on other models for water (see Ref.~\cite{Gallo2016} for discussion). Similar transitions were also investigated in models for carbon~\cite{ree09:prl}, silicon~\cite{sastry03:natm}, silica~\cite{poole05:ptrs}
and experimentally observed in phosphorus~\cite{Falconi2003}, triphenyl phosphite and n-butanol~\cite{Kurita2005}. Although much debated in the literature~\cite{Limmer2013a,Gallo2016}, recent experiments with mixtures of water and glycerol~\cite{Murata2013} and measurements of correlations functions using time-resolved optical Kerr effect (OKE) of supercooled water~\cite{Taschin2013} favor the SCP hypothesis. \\
\indent The debate would be further benefited if simple toy-models with waterlike behavior could be found and analyzed, especially if they could be probed experimentally in the neighborhood of the hypothetic phase transition. In this paper we show that the Bose-Hubbard model provides one such platform. \\
\indent In the following, the Bose-Hubbard model is investigated in the so-called ``atomic limit"~\cite{Hubbard1963ElectronBands} of vanishingly small hopping amplitude. Despite the simplicity of the model in this regime, a rich water phenomenology emerges. It is important to stress that an ``authentic" waterlike behavior could only be observed on the Bose-Hubbard model in the $NPT$ ensemble, and this partially explains why it remained unnoticed so far: theoretical and experimental communities working with this system are commonly using $\mu VT$ ensemble~\cite{Ho2010}, where no maximum of density can be found. \\
\indent Our proposal is supported by previous investigations which established a connection between ground state phase transitions (GSPT) and waterlike anomalies in the context of classical lattice and off-lattice models of fluids in one dimension~\cite{barbosa11jcp, Barbosa2013, barbosa:jcp2015, Rizzatti2018}. As in the SCP hypothesis, the model does present (ground state) critical behavior, although lacking finite temperature phase transitions.\\
\indent This paper is organized as follows: the Bose-Hubbard model and its ground state in the atomic limit are analyzed in section~\ref{sec:bh}, the grand canonical partition function and relevant thermodynamic quantities are calculated on section~\ref{sec:thermo}, with the detailed expressions for pressure and chemical potential left for the Appendix. Our results are discussed on section~\ref{sec:results} and the final remarks made on section~\ref{sec:final}. 

\section{The Bose-Hubbard Model and its grand canonical ground state\label{sec:bh}}

\indent On its simplest realization, the Bose-Hubbard model consists of a lattice whose sites are empty or occupied by a certain number of particles and its hamiltonian presents terms for hopping ($J>0$), chemical potential ($\mu$), and the on site interaction disfavoring multiple occupation on the same site ($U>0$). Creation and annihilation operators are defined as usual with symbols $\hat a_i^\dag$ and $\hat a_i$ and the number operator on site $i$ is $\hat n_i = \hat a_i^\dag \hat a_i $. With these definitions this hamiltonian becomes~\cite{fisher1989:prb}:
\be
\hat H   =   -\sum_{ \langle i,j \rangle} J \hat a_i^\dag \hat a_j +\sum_i \frac{U}{2} \hat n_i ( \hat n_i-1) - \sum_i \mu \hat n_i , \label{eq:qbh}
\ee	
where the first sum is over all pairs of nearest neighbor sites and the others involve all $L$ sites. \\
\indent Here we analyze the atomic limit by setting $J=0$. With this choice, tunneling between different sites is forbidden and the superfluid phase, which is composed by particles in a delocalized state, does not exist. While from the experimental point of view it would be equivalent to a very strong trapping field, from the theoretical perspective it allows us to maintain waterlike anomalous properties without dealing with the more complex quantum phenomenology of the superfluid phase.\\
\indent The hamiltonian becomes a sum of single-site hamiltonians $\hat H_i$, which can be diagonalized by using the number operators eigenvectors $\hat n_i \ket{n_i} = n_i \ket{n_i}$. Hence the energy eigenvalue of a single site with occupation $n_i=n$ becomes
\be 
  \epsilon_n = \frac{U}{2} n (n -1 ) -\mu n.
\ee
Since lattice sites are \textit{distinguishable}, quantum statistics end up identical to Boltzmann statistics~\cite[p. 17]{sachdev}. For this reason, and for the benefit of a broader audience, classical statistical mechanics is employed from this point throughout this article.\\ 
\indent We proceed by investigating the GSPT. At $T=0$ and a given $\mu$, the grand canonical free energy $\Phi=V\phi$ (volume $V=v_0 L$, with $v_0$ defining the lattice cell volume) is simply the result of the minimization procedure $\phi(T=0,\mu)=\min \limits_{n} \epsilon_n $. Therefore $\phi(T=0,\mu)=\epsilon_n$ for $n$ satisfying $(n-1)U<\mu<nU$. This implies that GSPT occur whenever the chemical potential hits an integer value of the on site interaction, where a coexistence between successive occupation states $n$ and $n+1$, called Mott Insulators, takes place. This analysis yields the critical chemical potentials  $\mu_n=nU$ and the corresponding critical pressures $P_n v_0 = n (n+1)\frac{U}{2}$.\\ 
\indent Calculating the densities that are observed in the GSPT at fixed chemical potential in the $\mu VT$ ensemble is simple and requires assuming that states $n$ and $n+1$ are equal \textit{a priori}. The result is $v_0\rho_n = n+1/2$ and will not be the same observed at fixed pressures in the $NPT$ ensemble, since  the pressure is a non differentiable function of $\mu$ at the GSPT. But these numbers can be obtained exactly within a two states description as will be explained in the Appendix~\ref{sec:two-states}.

\section{Thermodynamics\label{sec:thermo}}

\begin{figure}[ht]
\begin{center}
\includegraphics[scale=1]{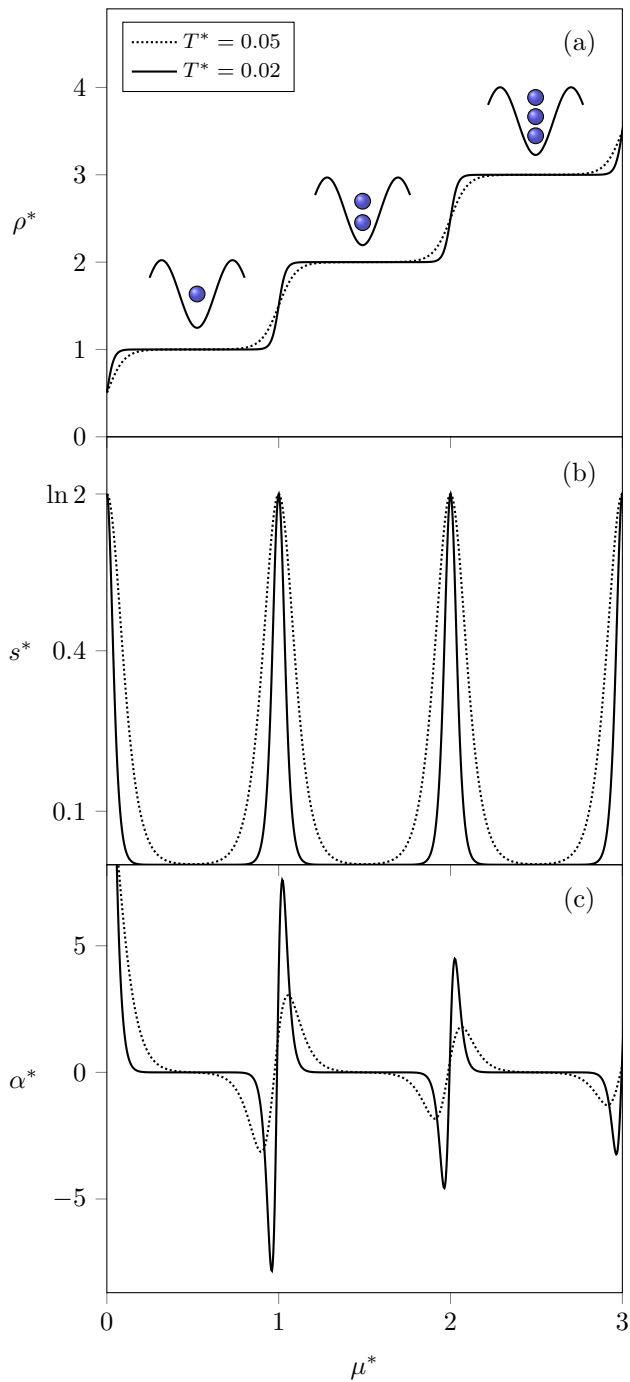}
\end{center} 
\caption{\label{fig:data-mu} (a) Density, (b) entropy, and (c) thermal expansion coefficient as a function of the chemical potential at fixed temperatures.}
\end{figure}

\indent The grand canonical partition function of the system can be expressed as:
\be
\Xi (T,V,\mu) = \left ( \sum_{n=0}^\infty  e^{-\beta \epsilon_n}   \right ) ^L,
\ee
where $\beta = 1/k_B T$, with $T$ being the temperature and $k_B$ the Boltzmann constant. Considering that
$\Xi = e^{-\beta \Phi}$ 
the fundamental relation for the grand thermodynamic potential $\Phi$ becomes:
\be
\Phi (T, V, \mu) = -k_B T L  \ln \left [ \sum_{n=0}^\infty e^{-\beta \epsilon_n} \right ].
\ee
Pressure can be obtained using $  \Phi = - PV $ and one can calculate density and entropy per site, employing the standard expressions: 
\be
\rho (T,\mu)  = \frac{N}{V} = -\frac{1}{V} \left ( \frac{\partial \Phi}{\partial \mu} \right)_{T},\label{eq:density}
\ee
and
\be
s (T,\mu) =  \frac{S}{V} = -\frac{1}{V} \left ( \frac{\partial \Phi }{\partial T} \right )_{\mu}.\label{eq:entropy}
\ee

\indent For the purpose of comparing this work with experimental realizations of the Bose-Hubbard model, it will be important to write the thermal expansion coefficient in terms of appropriate variables. Through a Jacobian transformation~\cite[p.~364]{Salinas:introduction} one obtains:
\begin{subequations}
\bea
\alpha & = & \frac{1}{V} \left ( \frac{\partial V }{ \partial T} \right )_{P,N} =  -\frac{1}{\rho} \left ( \frac{\partial \rho }{ \partial T} \right )_{P}  \label{eq:alpha} \\
            & = & \alpha_{\mu} +  \frac{1}{\rho} \left ( \frac{\partial \rho}{\partial \mu} \right )_T \dfrac{\left ( \dfrac{\partial \Phi}{\partial T} \right )_{\mu}}{\left ( \dfrac{\partial \Phi }{\partial \mu} \right )_{T}} \label{eq:alpha-b},
\eea
\end{subequations}
where $\alpha_{\mu}$ was defined as:
\be 
\alpha_{\mu} (T,\mu) = - \frac{1}{\rho} \left ( \dfrac{\partial \rho}{\partial T} \right )_{\mu}.\label{eq:alpha-mu}
\ee

\indent Expressions~(\ref{eq:density})-(\ref{eq:alpha-mu}) will be calculated in the $\mu VT$ ensemble and converted to the $NPT$ ensemble whenever necessary. 

\section{Results and discussions\label{sec:results}}

\indent Before proceeding let us note that variables are reduced in terms of $U$, $v_0$ and $k_B$, as $T^* = k_B T / U$, $\mu^* = \mu/U$ and $P^* = P v_0 /U$. Our analysis starts by comparing density, entropy and thermal expansion $\alpha$ as a function of chemical potential at fixed temperature (Fig.~\ref{fig:data-mu}). In this figure $\mu$ was used in the $x$-axis to facilitate comparison with theoretical/experimental data in the literature. Also note that $\alpha$ is the same used in the fluid literature, eq.~(\ref{eq:alpha}),  and was calculated from~(\ref{eq:alpha-b}). As it is well known, one-dimensional systems containing only short range interactions can only undergo phase transitions (characterized by discontinuities in the thermodynamic functions) at zero temperature \cite{VanHove1950}. Fig.~\ref{fig:data-mu} (a) shows that the density is highly sensitive to changes in the chemical potential around $\mu^*_n = n$, for integer $n$, and that this response becomes sharper at lower temperatures, approaching true phase transition discontinuities in the $T\rightarrow 0$ limit. This confirms that $\mu^*_n$ are indeed the critical values of the zero-temperature GSPT. \\
\indent On Fig.~\ref{fig:data-mu} (b) entropy is shown to develop maximum values exactly at the critical chemical potentials $\mu^*_n$. As temperature decreases entropy goes to zero except at the transition points, where it becomes sharper and turn into a residual entropy in the limit $T \rightarrow 0$. 
Note that the maximum equals $s^*_n = \ln 2$ which is expected for a two state mixture. From the Maxwell relation 
\be
\left ( \frac{\partial S}{\partial P} \right )_T = - \left ( \frac{\partial V}{\partial T} \right )_P  = - V \alpha,
\ee
it follows that $\alpha$ is negative (positive) whenever entropy increases (decreases) with pressure~\footnote{chemical potential monotonically increases with pressure}. Thus, an entropy maximum introduces an oscillation in thermal expansion $\alpha$, with its amplitude increasing as temperature is lowered according to Fig.~\ref{fig:data-mu} (c). The oscillations evolve to a peculiar double divergence with $\alpha \rightarrow -\infty$ ($+\infty$) as $\mu \rightarrow \mu^-_n$ ($\mu^+_n$). Indeed, such mechanism establishes a quite general connection between GSPT, residual entropy and density anomaly. The multiple configurations remaining from each critical point produce a macroscopic zero point entropy. When temperature is raised, the possibility of the system accessing these states can induce an anomalous behavior depending on the chosen external fields.   

\begin{figure}[t]
\begin{center}
\includegraphics[scale=1]{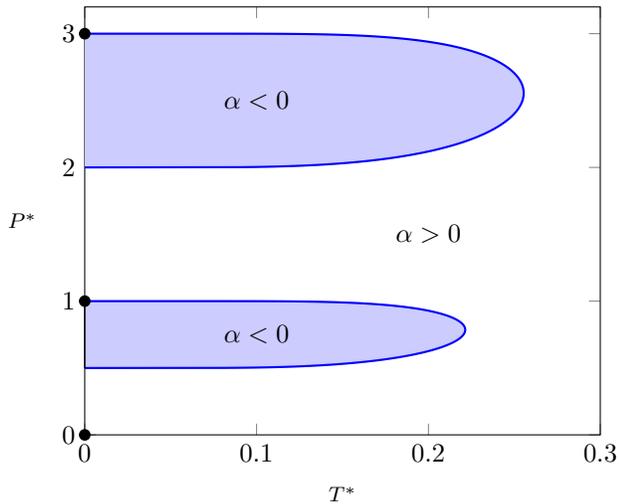}
\end{center}
\caption{\label{fig:tmd} Pressure vs. temperature  phase diagram with GSPT marked with filled circles and continuous lines representing the TMD. The anomalous states are represented by the filled  areas.}
\end{figure}

\begin{figure}
\begin{center}
\includegraphics[scale=1]{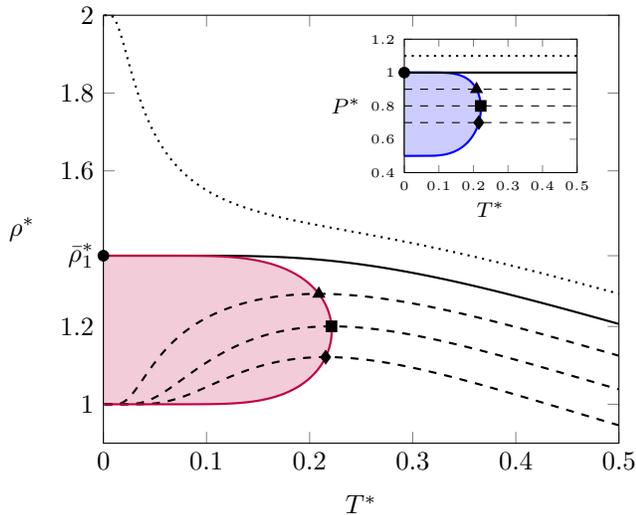}
\end{center}
\caption{\label{fig:density-p} Density as a function of temperature for fixed pressures. Density increases with temperature (region filled in purple) and presents a maximum (highlighted points over the continuous purple curve) for pressures slightly below (dashed lines) the critical pressure $P^*_1=1$ (continuous black line). Density decreases monotonically with temperature for pressures above  (dotted line) the critical value. The inset contains the $P \times T$ phase diagram featuring a TMD line and the pressures chosen.}
\end{figure}

\indent Next we discuss the emergence of TMD lines in the phase diagram of the model. On Fig.~\ref{fig:tmd} their~\textit{loci}, corresponding to $\alpha=0$, are shown in a range of pressures covering two regions where density increases with temperature ($\alpha<0$). As in our previous studies~\cite{barbosa:jcp2015,barbosa11jcp}, TMD lines are emanating from GSPT (filled circles) and draws a curve enclosing a region of the phase diagram starting and ending at $T=0$. The endpoint of these lines can be obtained by analyzing enthalpy variations for adding or excluding a particle in the system. Even though we have chosen to show two TMD lines starting from transitions at $P^*_1 = 1$ and $3$, the model exhibits an infinite number of GSPT and also an infinite number of regions in the $P \times T$ phase diagram where $\alpha < 0$.\\
\indent A more detailed view on the density behavior is presented in Fig.~\ref{fig:density-p}, where it is plotted against temperature at pressures slightly above, below and equal to the critical value $P^*_1 = 1$. It is interesting to observe that density increases with temperature below $P^*_1$, reaching a maximum value and then decreasing again, while above $P^*_1$ density decreases, as in a normal fluid. Exactly at $P^*_1$ density reaches a fixed value at about the same temperature where the TMD line becomes horizontal in the $P \times T$ phase diagram (see the inset of Fig.~\ref{fig:density-p}). It is possible to calculate this value within a low temperature, two-states expansion (see Appendix), resulting in the polynomial:
\be
 (1-2\delta_n)^{n+1} = 2(1+2\delta_n)^n \label{eq:density-maximum-k}
\ee
where $\delta_n = \bar{\rho}^*_n - \rho^*_n$, with $\bar{\rho}^*_n$ ($\rho^*_n = n + 1/2$) being the critical density at fixed pressure (fixed chemical potential) for the $n$-th transition. From this it is possible to find $\bar{\rho}^*_1$, the critical density at constant pressure for $n=1$, as
\be
\bar{\rho}^*_1 = \frac{5 - \sqrt{5}}{2} \approx 1.381966. \label{eq:density-maximum-k=1}
\ee
Accordingly, the critical densities obtained from~Eq.(\ref{eq:density-maximum-k}) are indeed relevant as they \textit{predict the maximum densities found along the TMD lines} emanating from GSPT at critical pressures $P^*_n = n(n+1)/2$.

\begin{figure}
\begin{center}
\includegraphics[scale=1]{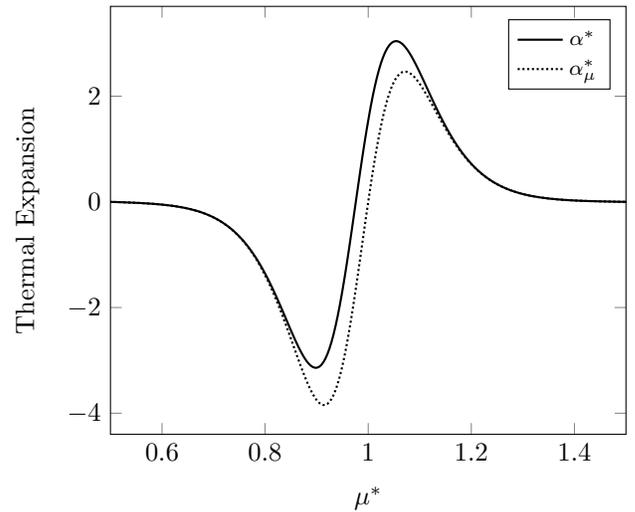}
\end{center}
\caption{ \label{fig:alpha} Thermal expansion coefficients at fixed pressure and chemical potential, $\alpha^*$ and $\alpha^*_{\mu}$, as a function of chemical potential at temperature $T^*=0.05$. At low temperature the behavior of both coefficients are similar (see text).}
\end{figure}

\indent Next, let us compare the low temperature aspects of $\alpha$ and $\alpha_{\mu}$ by rewriting Eq.~(\ref{eq:alpha-b}) as:
\be
\rho (\alpha - \alpha_{\mu}) = \frac{ s }{ \rho } \left ( \frac{\partial \rho}{\partial \mu} \right )_T. \nonumber
\ee
At small temperatures, it follows from the r.h.s. of this expression that  $\alpha \approx \alpha_{\mu}$ for $\mu \neq \mu_n$ since $\lim\limits_{T \to 0} s \rightarrow 0$. Consequently, $\alpha$ and $\alpha_{\mu}$ are resembling functions at low temperatures, and $\alpha_{\mu} < 0$ can be used to infer a waterlike behavior in the $NPT$. As shown in Fig.~\ref{fig:alpha}, near the ground state phase transition between fluid phases with $n=1$ and $n=2$ particles at $\mu^*_1 = 1$, $\alpha_\mu$ presents an oscillation similar to $\alpha$, this being a signature of the proximity to the GSPT and waterlike behavior~\cite{barbosa:jcp2015}.

\section{Conclusion\label{sec:final}}

\indent In this work waterlike volumetric anomalies were observed in the Bose-Hubbard model by considering the so-called atomic limit, where the confining field is sufficiently intense such that particle hopping across the lattice is strongly suppressed. Ground state analysis reproduced the expected phase transitions between Mott insulators with different fillings.\\
\indent The grand canonical partition function was calculated and density, entropy and thermal expansion coefficient (at fixed pressure) were shown to behave anomalously in certain regions of the phase diagram. It was found that TMD lines were emerging from GSPT, being associated to residual entropy occurring on these transitions. This points towards a connection between phase transitions, residual entropies and density anomalies, and how these effects come together to produce an oscillatory thermal expansion coefficient, a hypothesis that was explored in previous works~\cite{barbosa11jcp, Barbosa2013, barbosa:jcp2015, Rizzatti2018}. It was demonstrated that at low temperatures the thermal expansion coefficient $\alpha$ is approximately equal to $\alpha_\mu$, and that the oscillatory behavior can be observed in both. This fact should be helpful, as it allows to identify regions where waterlike anomalies are expected to happen in the $NPT$ ensemble while looking at the behavior of $\alpha_\mu$ in $\mu V T$ ensemble at low temperatures.\\
\indent The fact that the Bose-Hubbard model presents waterlike behavior opens a new range of possibilities to theoretically and experimentally investigate the waterlike phenomenology and its relation to phase transitions. Of obvious interest would be to test the proposed theoretical scenarios using already available cold atom realizations of the Bose-Hubbard model in the atomic limit.\\
\indent We acknowledge useful discussions with Marcia Barbosa. This work has been supported by the Brazilian funding agencies CNPq and CAPES. \\

\appendix*

\section{Two-states approximation and the critical densities\label{sec:two-states}}

\indent Near the GSPT between configurations of occupation numbers $n$ and $n+1$, the grand canonical free energy can be approximated by
\be
\Phi \approx -\frac{1}{\beta} \ln\left ( e^{-\beta \epsilon_n} + e^{-\beta \epsilon_{n+1}}  \right )^L,
\ee
from which we calculate pressure as
\bea
P v_0 & \approx  & - \frac{ \epsilon_{n}+\epsilon_{n+1} }{2} \nonumber  \\
      &          & + \frac{1}{\beta} \ln \left \{  2 \cosh \left [ \frac{\beta ( \epsilon_{n}-\epsilon_{n+1} )}{2}  \right] \right \}.
\eea
Now we define $\Delta P  =  P - P_n$ and $\Delta \mu  =   \mu - \mu_n$ to rewrite
\be
\Delta P v_0 = \left ( n+\frac{1}{2} \right ) \Delta \mu + \frac{1}{\beta}  \ln \left [  2 \cosh \left( \frac{\beta\Delta \mu}{2}  \right) \right ], 
\label{eq:p-mu}
\ee
and calculate
\be
\rho v_0 =  \left ( n+\frac{1}{2} \right ) + \tanh \left ( \frac{\beta \Delta \mu}{2} \right). \label{eq:mu-rho}
\ee
By inverting Eq.~(\ref{eq:mu-rho}) it is possible to obtain
\be
e^{\beta \Delta P^*} (1-2\delta_n)^{n+1} = 2(1+2\delta_n)^n,
\ee
with $\delta_n =  \bar{\rho}^*_n - \rho^*_n$ as defined above.  At the critical pressure, $\Delta P^*=0$ and Eq.~(\ref{eq:density-maximum-k}) is recovered. The case $n=1$ leads to the second order polynomial
\be
4\delta^2_1 - 8\delta_1 -1=0,
\ee
whose physically viable solution is $\delta_1 = (2-\sqrt{5})/2$, resulting in $\bar{\rho}^*_1= (5 - \sqrt[]{5})/2$, as discussed in the section~\ref{sec:results}. The values of the critical densities for arbitrary $n$ can be calculated numerically from equation~(\ref{eq:density-maximum-k}). These solutions have the property $\lim\limits_{n \rightarrow \infty} \delta_n = 0$, meaning that in this limit critical densities become identical when calculated at fixed $\mu$ and fixed $P$.

\end{document}